\input{aipcheck}

\documentclass[
    ,final            
  ]
  {aipproc}

\layoutstyle{6x9}

\begin{document}

\title{On the release of binding energy and accretion power
in core collapse-like environments}


\author{Aristotle Socrates}{
  address={Department of Astrophysical Sciences,
Peyton Hall-Ivy Lane, 
 Princeton University\\
Princeton, NJ 08544}
}

\author{Enrico Ramirez-Ruiz}{
  address={Department of Astronomy, University of California\\
Santa Cruz, CA 93106}}

\begin{abstract}
All accretion models of gamma-ray bursts share a common assumption:
accretion power and gravitational binding energy is released and 
then dissipated locally, with the mass of its origin.  This is
equivalent to the Shakura-Sunyaev 1973 (SS73) prescription for the
dissipation of accretion power and subsequent conversion 
into radiate output.  Since their seminal paper, broadband 
observations of quasars and black hole X-ray binaries
insist that the SS73 prescription cannot wholly describe
their behavior.  In particular, optically thick black
hole accretion flows are almost 
universally accompanied by coronae whose relative power by far 
exceeds anything seen in studies of stellar chromospheric and
coronal activity.  In this note, we briefly discuss the possible 
repercussions of freeing accretion models of GRBs from the 
SS73 prescription.  Our main conclusion is that the efficiency 
of converting gravitational binding energy into a GRB power
can be increased by an order of magnitude or more.    
\end{abstract}

\maketitle


\section{Introduction}

A parcel of matter in a Keplerian orbit about a  
dominant central source of gravity has the potential to release
some fraction of $\frac{GM}{Rc^2}$, multiplied
by its rest mass energy via the act of accretion.  The value 
of this fraction, often referred to as the accretion efficiency 
$\epsilon$, depends upon the compactness of the massive 
central object.  For a solar-type star $\epsilon\sim 10^{-6}$,
while for a white dwarf $\epsilon\sim 10^{-3}$.  Neutron stars and
black holes -- the most compact objects in the Universe --
possess the largest values for the accretion efficiency,
with $\epsilon\sim 10^{-1}$.  

The total energy release from the act of accretion $\Delta E$ is
therefore equal to the total mass accreted $\Delta M$ multiplied by
$\epsilon$ i.e., $\Delta E=\epsilon\Delta M$.  It follows that the
total accretion power $L_{\rm acc}$ is given by $L{_{\rm acc}}=\Delta
E/\Delta\, t$, where $\Delta\, t$ is the characteristic accretion
time.  These simple considerations, along with realizing that
gamma-ray bursts (GRBs) trigger at cosmological distances,
provided workers to explain their large and rapid energy release
by the accretion of $\Delta M\sim M_{\odot}$ of matter in
time $\Delta t\sim$ a few seconds (Goodman et al. 1987;
Narayan et al 1992; Woosley 1993).  The basic idea is that
somehow, accretion power is partially converted into outgoing 
mechanical power of electronically interacting particles such 
as pairs and baryons (the so-called ``fireball'' model)
or coherent photon states (so-called ``Poynting flux dominated''
jets). 

We have yet to specify in which manner, or through which 
channels, gravitational binding energy is ${\bf ( A)}$ initially
randomized ${\bf (B)}$ how that randomized gravitational energy is 
converted into particle energy  
${\bf (C)}$ and finally, the manner in which particle energy
is then converted into radiation which ultimately removes
binding energy from the system.  

In the last 15 years, a tentative consensus -- despite 
the lack of any concrete observational evidence -- has been 
reached with respect ${\bf (A)}$ (Balbus \& Hawley 1991; 1998). 
However, knowledge of ${\bf (A)}$ only informs us 
of the value of $\Delta t$ and $L_{_{\rm acc}}$ for a given 
energy reservoir $\Delta E$.  Ultimately, the physical processes
that determine ${\bf (B)}$ and ${\bf (C)}$ allow for theorists to infer 
the properties of the fraction of $\Delta E$ that goes into 
powering the highly relativistic (Lorentz factor $\Gamma\sim 
E/M\sim 100$) 
GRB outflow.   

Admittedly, our conversation has been abstract.  In the next few 
sections we become more concrete as we discuss the the basic features 
of hyper-Eddington accretion disks (HEDs) and then closely examine 
they key idea that all workers have assumed when considering their 
structure and applicability to GRBs.

\begin{figure}
  \includegraphics[height=.3\textheight]{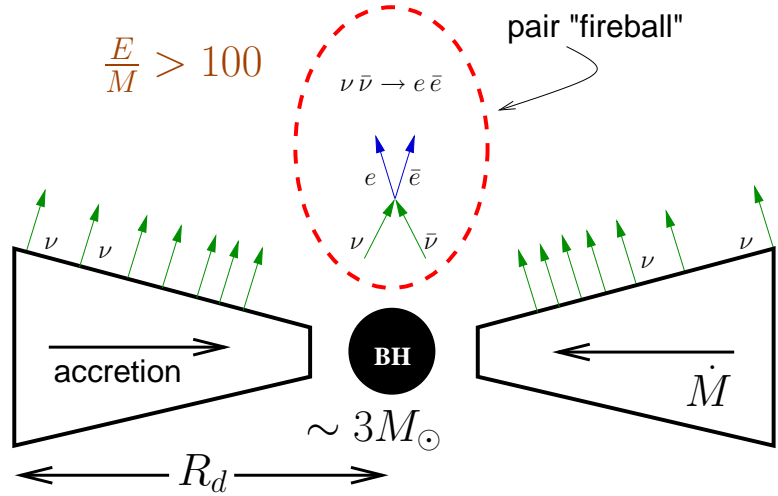}
\label{f: model}
  \caption{Accretion + fireball model of gamma ray bursts. The 
basic ingredients are a stellar mass black hole and a debris torus
of non-negligible mass i.e., the mass of the disk is a 
significant fraction of a solar mass.  The tidal radius of the disk 
$R_d\sim$$ 10^3$ gravitational radii $R_g$, which is roughly the 
radius of a stellar mass object supported by relativistic electron
degeneracy pressure (e.g, a white dwarf or burnt-out iron core).  
Under these rather
extreme conditions, it is possible to generate $\epsilon
\Delta M\sim$ a few $\times 10^{53}$ ergs worth or accretion 
energy, where $\Delta M$ is given by the mass of debris torus.}
\end{figure}

\section{Hyper-Eddington Disks as the central engines of GRBs
and the Shakura-Sunyaev (1973) Prescription.}

Only a few plausible astrophysical scenarios can lead to a HED, 
as depicted in Figure \ref{f: model}.  In recent years, the most 
popular progenitor model has been the ``collapsar'' model (Woosley 
1993; MacFadyen \& Woosley 1999), which is involves the progenitor
of a core-collapse supernova (SN) whose degenerate iron core possesses
an appreciable amount of angular momentum.  Rather than forming a 
proto-neutron star, which are thought to power ``normal'' core-collapse
SNe, some combination of a stellar mass BH and a massive debris torus
is formed.  That is, matter initially lying near the progenitor's axis
rotation axis falls towards the systems gravitational center, thus 
making a prompt contribution to the formation of a BH.  Whereas, 
matter originating near the progenitor's equator forms a disk whose 
tidal radius is determined by initial specific angular momentum of the 
debris before the onset of collapse.

The viscous time of the disk $\Delta\, t$ is given by
\begin{equation}
\Delta\, t \sim  R^2_d/\nu,
\end{equation}
where 
 along with it's 
mass, determines the accretion rate $\dot{M}=\Delta M/\Delta t$.
The Shakura-Sunyaev (1973; hereafter SS73)  prescription 
for the kinematic viscosity $\nu$ reads
\begin{equation}
\nu\sim \alpha c_s\,H,
\label{stress}
\end{equation}
where $\alpha$, $c_s$ and $H$ is the SS73 ``$\alpha$- parameter'',
the sound speed of the disk and the disk half-thickness,
respectively.  The parameter $\alpha$ is equal to the ratio of the 
accretion stress, $\tau_{R\phi}$ to the disk pressure at the 
midplane $P$.  If the tidal radius
$R_d$ corresponds to $\sim$ a few $\times 10^3\, R_g$  for a 
stellar mass BH and the disk is only marginally thin such 
that $H/R\leq 1$, which implies that $c_s\leq R\Omega$, then
$\Delta t\sim$ a few seconds.  If the mass fuel supply $\Delta M\sim
M_{\odot}$, it follows that the accretion power takes on formidable
values upwards of $L_{\rm acc}\sim$ a few 
$\times 10^{52}{\rm erg\,s^{-1}}$, five orders of magnitude
larger than the brightest quasars.  

In order to determine the value of $\Delta t$, the value of $\alpha$
had to be specified.  For almost two decades, theorists typically 
assumed that 
$\alpha\sim 10^{-1}-10^{-2}$, for {\it almost} no particular reason.
\footnote{Timing studies of cataclysmic variables indicate that
$\alpha\sim 0.1$ (Pringle 1981).}  Now theorists believe MHD
turbulence is the source of the accretion stress 
$\tau_{R\phi}$ (Balbus \& Hawley 1991), with calculated values
of $\tau_{R\phi}/P\sim 10^{-1}-10^{-2}$, in a satisfying act of 
validation.  With this, we have a simple, plausible
and therefore attractive, working hypothesis for 
the mechanism responsible for ${\bf (A)}$
i.e., the initial randomization of gravitational 
binding energy.

In order to determine the structure of a HED, a condition that
enforces conservation of energy, or radiative equilibrium, must
be specified.  At any given radius, the input of energy is 
roughly quantified by the rate viscous dissipation, while 
energy losses result from some combination of the inward advection
of heat and radiative output.

The accretion rate ${\dot M}$ and the accretion power 
$L_{\rm acc}$ are enormous because the reservoir of mass is
proportionally large, hence the term {\it hyper-Eddington} 
accretion disk.  Consequently, the Thomson optical depth for
these HEDs are so high that the photon diffusion time greatly 
exceeds the inward advection time and therefore, the emission 
of photons cannot contribute to cooling the disk.\footnote
{It's straightforward to show that the Thomson optical 
depth $\tau_{\rm es}\geq {L_{\rm acc}}/{L_{\rm Edd}}\geq
{L_{\rm GRB}}/{L_{\rm Edd}}$, where $L_{\rm Edd}$ is the 
Eddington luminosity and $L_{\rm GRB}$ is the luminosity 
of the GRB.  Therefore, $\tau_{\rm es}\geq 10^{12}$, which 
implies that the photon diffusion speed $c/\tau_{\rm es}$
is so small that photon diffusion can safely be ignored.}  
Furthermore, 
the enormous accretion power $L_{\rm acc}$ also implies that the 
characteristic disk temperatures are large as well.  In fact, 
near the hole -- where most of $L_{\rm acc}$ is generated --
the temperatures and densities of the flow resemble that of a 
proto-neutron star's surface i.e., an object that can 
can generate {\it neutrinos} in copious amounts via the 
reactions 
\begin{equation}
e+p\rightarrow n+\nu_e\,\,\,\,\,{\rm AND}\,\,\,\,\,\,
e^++n\rightarrow p+\bar{\nu_e}.
\end{equation}

If a significant amount $L_{\rm acc}$ is liberated in the form 
of neutrino radiation, then the condition for radiative equilibrium 
is approximately given by
\begin{equation}
H\,\tau_{R\phi}\frac{d\Omega}{d{\rm ln}R} \sim  \sigma T^4_{\rm eff}
\label{rad1}
\end{equation}
or in words, 
\begin{equation}
{\rm ACCRETION\,\,\,\, POWER}  \sim  {\rm THERMAL\,\,
EMISSION}.\nonumber
\label{rad2}
\end{equation}

The notion encapsulated by eqs. (\ref{rad1}) and (\ref{rad2}) is
the same dissipation prescription adopted by SS73.  The 
supposition that
accretion power is matched by thermal radiation implies some
knowledge of the mechanisms that account for ${\bf (B)}$
and ${\bf (C)}$.  That is, randomized binding energy
ultimately heats particles that can easily radiate their energy 
on an accretion time $\Delta t$.  Furthermore, the 
spatial {\it location} of the particles that receive the binding 
energy must roughly coincide with the particles (baryons) that
serve as the source of gravitational and accretion 
energy.  In doing so, the absorption optical depth under which 
the accretion power is released into heat is relatively large,
as long as the absorption opacity increases with density,
and the resultant outgoing radiation therefore approaches  
a black body.  This is the meaning of {\it local} dissipation 
of binding energy i.e., the overwhelming majority of the 
binding energy is dissipated in the overwhelming 
majority of the matter.    

The Shakura-Sunyaev (1973) prescription for accretion, 
encapsulated by eqs. (\ref{stress}) and (\ref{rad1}), 
has profound consequences for the HED-powered 
fireball model of GRBs, outlined in Figure
\ref{f: model}.  Woosley (1993) took note that the 
most straightforward way of energizing a pair-rich 
relativistic plasma along the flow's axis of symmetry 
is the annihilation of disk neutrinos into 
$ee^+$ pairs.  The neutrino annihilation 
deposition rate is given by (neglecting geometric factors)
\begin{equation}
Q^+_{\nu ,{\bar\nu}}=\sigma_{0}\frac{L^2_{\nu}}{A^2}
\left<E_{\nu}\right>.
\label{pairs}
\end{equation}
Definitions of the various symbols in the above 
expression are given in Ramirez-Ruiz and Socrates (2005).
For a fixed value of $L_{\rm acc}\sim L_{\nu}$,
the basic thermodynamics of radiating sources informs us
that the quantity $L_{\nu}\left<E_{\nu} \right>$, which 
is $\propto Q^+_{\nu ,{\bar\nu}}$, assumes it's smallest 
value as the source approaches the black body limit.  As is 
well known, black body radiation is ``efficient'' because it
releases a given amount of radiant power at the lowest 
possible temperature or energy per quanta.  The physical 
reason why the deposition rate takes on the functional 
dependence of eq. (\ref{pairs}) is because the neutrino 
cross section increases with the square of the neutrino 
energy.  In short, utilization of the Shakura-Sunyaev
prescription, which implicitly assumes local dissipation,
is the least efficient mode of gravitational energy 
release in terms of energizing a GRB pair fireball
with a HED disk. 

Interestingly, stellar mass BH accretion in X-ray binaries (BHXRBs)
and quasars i.e, black hole accretion that we can actually observe,
does not heed the Shakura-Sunyaev prescription.  Even in the
``high-soft'' state of BHXRBs or quasars, sources in which a cool
optically thick thermal-emitting disk dominates the {\it photon}
spectral energy distribution (SED), a significant fraction ($\sim
10\%-50\%$) of the accretion power $L_{\rm acc}$ is released in the
form of hard Comptonized X-rays.  The source of Compton power is
widely believed to originate from a hot diffuse corona, adjacent to
the cool optically thick disk.  In other words, a disproportionately
small amount of matter is responsible for a disproportionately large
amount of radiative energy release, the implication being that the
assumption of local dissipation is invalid and that {\it black hole
accretion flows observed in Nature dissipate their gravitational
binding energy non-locally.}  The net effect on the SED is that the
outgoing radiation field may display significant distortions from a
thermal black body spectrum, implying that the average energy per
quanta in the disk + corona case is larger in comparison to the purely
thermal disk situation.

\begin{figure}
\label{SED}
  \includegraphics[height=.3\textheight]{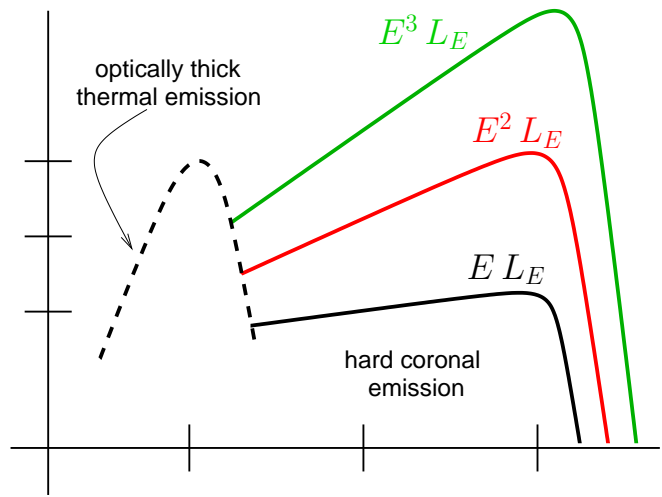}
  \caption{Possible neutrino spectral energy distribution from 
a hyper-Eddington disk.  The $y-$axis is the logarithm of the 
spectral energy distribution and the $x-$axis is the logarithm of 
the emitted radiation's energy.}
\end{figure}

If we now lift the extreme condition thar the mode of energy release in 
HEDs are purely described by the Shakura-Sunyaev (1973) prescription,
then we may break the integral $L_{\nu}\left<E_{\nu}\right>$ into 
a soft component coming from a cool dense disk and the other 
from a hot diffuse corona
\begin{equation}
L_{\nu}\left<E_{\nu}\right>=L_{\nu}\left[\left(1-f\right)
\left<E_{\nu}\right>_s + f\left<E_{\nu}\right>_c\right]
\end{equation}
where $f$ is the fraction of power released in a
corona and $\left<E_{\nu}\right>_s$ and 
$\left<E_{\nu}\right>_c$ is the average neutrino 
energy emitted from the disk and corona, respectively.
So, if $f\sim 1\%$ of $L_{\rm acc}$ is released in a
corona with a temperature that is $1/f\sim 100$ times larger
than the disk temperature, then the energy 
deposition rate resulting from pair annihilation from both 
regions of the flow are equal.  Clearly, the possibility 
that some fraction of $L_{\rm acc}$ in HEDs is released
non-locally in a corona can have profound effects 
on the net energetics of the GRB explosion. Consider 
Figure \ref{SED}.  The black curve is the luminosity 
SED, which indicates the amount of power radiated 
per decade in energy, while the red and green curve 
represents the normalized (with respect to the 
soft disk contribution) deposition rate for the 
pair-annihilation process (relevant for HEDs as
GRB engine) and capture of electron-type neutrinos
(relevant for proto-neutron stars as SN engines), 
respectively.   

The basic question of whether or not a HED scenario can 
energize a GRB fireball raises an even more fundamental 
question in a broader astrophysical sense.  That is, what 
are the physical mechanisms at work in determining 
the ${\bf (A)}$${\bf (B)}$${\bf (C)}$s of accretion?  If 
neutrino-emitting hyper-Eddington black hole accretion flows
work anything at all like their photon-emitting sub-Eddington 
counterparts, then the possibility that a significant fraction
of the accretion power is released through a corona can no longer
be ignored.

\begin{theacknowledgments}
The authors thank the referees from the core-collapse community
for their reports on the manuscript astro-ph/0504257.
Responding to their comments has made the friendship between 
the authors a closer one.
  
\end{theacknowledgments}



\bibliographystyle{aipproc}   

\bibliography{sample}

\IfFileExists{\jobname.bbl}{}
 {\typeout{}
  \typeout{******************************************}
  \typeout{** Please run "bibtex \jobname" to optain}
  \typeout{** the bibliography and then re-run LaTeX}
  \typeout{** twice to fix the references!}
  \typeout{******************************************}
  \typeout{}
 }

\end{document}